
\magnification 1200
%

%
\font\eightrm=cmr8
\font\eighti=cmmi8
\font\eightsy=cmsy8
\font\eightbf=cmbx8
\font\eighttt=cmtt8
\font\eightit=cmti8
\font\eightsl=cmsl8
\font\sixrm=cmr6
\font\sixi=cmmi6
\font\sixsy=cmsy6
\font\sixbf=cmbx6
\catcode`@11
\newskip\ttglue
\font\grrm=cmbx10 scaled 1200

\def\eightpoint{\def\rm{\fam0\eightrm}
\textfont0=\eightrm \scriptfont0=\sixrm \scriptscriptfont0=\fiverm
\textfont1=\eighti \scriptfont1=\sixi \scriptscriptfont1=\fivei
\textfont2=\eightsy \scriptfont2=\sixsy \scriptscriptfont2=\fivesy
\textfont3=\tenex \scriptfont3=\tenex \scriptscriptfont3=\tenex
\textfont\itfam=\eightit \def\it{\fam\itfam\eightit}
\textfont\slfam=\eightsl \def\sl{\fam\slfam\eightsl}
\textfont\ttfam=\eighttt \def\tt{\fam\ttfam\eighttt}
\textfont\bffam=\eightbf
\scriptfont\bffam=\sixbf
\scriptscriptfont\bffam=\fivebf \def\bf{\fam\bffam\eightbf}
\tt \ttglue=.5em plus.25em minus.15em
\normalbaselineskip=6pt
\setbox\strutbox=\hbox{\vrule height7pt width0pt depth2pt}
\let\sc=\sixrm \let\big=\eightbig \normalbaselines\rm}
\newinsert\footins
\def\newfoot#1{\let\@sf\empty
  \ifhmode\edef\@sf{\spacefactor\the\spacefactor}\fi
  #1\@sf\vfootnote{#1}}
\def\vfootnote#1{\insert\footins\bgroup\eightpoint
  \interlinepenalty\interfootnotelinepenalty
  \splittopskip\ht\strutbox 
  \splitmaxdepth\dp\strutbox \floatingpenalty\@MM
  \leftskip\z@skip \rightskip\z@skip
  \textindent{#1}\footstrut\futurelet\next\fo@t}
\def\fo@t{\ifcat\bgroup\noexpand\next \let\next\f@@t
  \else\let\next\f@t\fi \next}
\def\f@@t{\bgroup\aftergroup\@foot\let\next}
\def\f@t#1{#1\@foot}
\def\@foot{\strut\egroup}
\def\footstrut{\vbox to\splittopskip{}}
\skip\footins=\bigskipamount 
\count\footins=1000 
\dimen\footins=8in 

\def\ref#1{$^{#1}$}
\def\flex{\raise 6pt\hbox{$\leftrightarrow $}\! \! \! \! \! \! }
\def\oversome#1{ \raise 8pt\hbox{$\scriptscriptstyle #1$}\! \! \! \! \! \! }

\newbox\bigstrutbox
\setbox\bigstrutbox=\hbox{\vrule height10pt depth5pt width0pt}
\def\bigstrut{\relax\ifmmode\copy\bigstrutbox\else\unhcopy\bigstrutbox\fi}
\def\refer[#1/#2]{ \item{#1} {{#2}} }
\def\rev<#1/#2/#3/#4>{{\it #1\/} {\bf#2}, {#3}({#4})}
\def\boxit#1{\vbox{\hrule\hbox{\vrule\kern3pt
\vbox{\kern3pt#1\kern3pt}\kern3pt\vrule}\hrule}}

\def\2figure#1#2#3#4{\vbox{ \hrule width#1truecm \hbox{\vrule height#2truecm
\hskip #1truecm
\vrule height#2truecm }\hrule width#1truecm \hbox{\vrule\vbox{\hsize #1truecm
\baselineskip=10pt
\noindent\strut#3}\vrule}\hrule width#1truecm
\hbox{\vrule\vbox{\hsize #1truecm
\baselineskip=10pt
\noindent\strut#4}\vrule}\hrule width#1truecm  }}
\def\3figure#1#2#3#4#5{\vbox{ \hrule width#1truecm \hbox{\vrule height#2truecm
\hskip #1truecm
\vrule height#2truecm }\hrule width#1truecm \hbox{\vrule\vbox{\hsize #1truecm
\baselineskip=10pt
\noindent\strut#3}\vrule}\hrule width#1truecm
 \hbox{\vrule\vbox{\hsize #1truecm
\baselineskip=10pt
\noindent\strut#4}\vrule}
\hrule width#1truecm \hbox{\vrule\vbox{\hsize #1truecm
\baselineskip=10pt
\noindent\strut#5}\vrule}\hrule width#1truecm  }}

\def\sqr#1#2{{\vcenter{\hrule height.#2pt
   \hbox{\vrule width.#2pt height#1pt \kern#1pt
    \vrule width.#2pt}
    \hrule height.#2pt}}}
\def\dal{\mathchoice{\sqr{6}{4}}{\sqr{5}{3}}{\sqr{5}3}{\sqr{4}3} \, }


\def\smin{\,\raise 0.06em \hbox{${\scriptstyle \in}$}\,}
\def\smsubset{\,\raise 0.06em \hbox{${\scriptstyle \subset}$}\,}

\def\Natural{\hbox{\hskip 1.5pt\hbox to 0pt{\hskip -2pt I\hss}N}}

\def\Rational{\hbox{\hbox to 0pt{\hskip 2.7pt \vrule height 6.5pt
                                  depth -0.2pt width 0.8pt \hss}Q}}
\def\Real{\hbox{\hskip 1.5pt\hbox to 0pt{\hskip -2pt I\hss}R}}
\def\Complex{\hbox{\hbox to 0pt{\hskip 2.7pt \vrule height 6.5pt
                                  depth -0.2pt width 0.8pt \hss}C}}


\def \qaz {\partial _ } \def \wsx {\varphi }
\nopagenumbers
\vfill\eject
\centerline {\grrm On the Dirac Quantization of Two Dimensional Gravity}
\vskip 1.5cm
\centerline {M.C.B. Abdalla\ref{1}, F.P.
Devecchi\newfoot{$^*$}{Present Address: Instituto de F\'{\i}sica,
Universidade Federal do Rio Grande do
Sul CP 15.051, Cep 91.500, Porto Alegre, R.S., Brazil}}
\vskip .5cm
\centerline { Instituto de F\'\i sica da Universidade Estadual Paulista, Julio
de Mesquita}
\centerline{ Rua Pamplona, 145, CEP 01405, S\~ao Paulo, Brazil}
\centerline{E. Abdalla}
\centerline { Instituto de F\'\i sica da Universidade de S\~ao Paulo}
\centerline { C.P. 20516, S\~ao Paulo, Brazil.}
\vskip 1.5cm
\centerline {\bf Abstract}
\vskip.4cm

\noindent We discuss the Dirac quantization of two dimensional gravity with
bosonic matter fields. After defining the extended Hamiltonian it is possible
to fix the gauge completely. The commutators can all be obtained in closed
form; nevertheless, the results are not particularly simple.

\vskip.5cm

\hfill Universidade de S\~ao Paulo\quad

\hfill IFUSP-preprint-1041\phantom{Paulo}\quad

\hfill March 1993\phantom{Paulo}\quad

\vfill

\vfill\eject

\pageno=1
\footline={\hss\tenrm\folio\hss}

\centerline{\bf Introduction}
\vskip .5cm

Two dimensional gravity has been studied in detail during the last few
years\ref{1,2}. Several important and clear results have been obtained for
Liouville theory\ref{3} as well as for the light cone gauge pure
gravity\ref{1,2}. Moreover correlation functions involving dressed operators
have been systematically computed\ref{4}. In this plethora of results, it is
however disappointing that very few methods can shed some light in the
canonical structure of higher dimensional gravitation theory, since two
dimensional
space-time has revealed strength as a theoretical laboratory for higher
dimensional (specially gauge) theories\ref{5}. Thus we shall follow
the works [6] and [7] using the canonical method in order to study two
dimensional  gravity.

We shall start with a brief discussion of pure two dimensional gravity. This is
a very simple case, since the number of constraints is too large, and the
canonical formalism fails to provide a non trivial result. However, working in
an analogous way as in ref.[6], it is possible to calculate some fundamental
quantities before the gauge fixing procedure.

To begin with we consider the two dimensional gravity Lagrangian
$$L=\sqrt {-g}\left(-{1\over 2}\varphi \dal \varphi -{\alpha \over 2}R\varphi
+ {\alpha ^2\over 2}\beta \right)\quad ,\eqno (1)$$
where $R$ is the scalar curvature, $\beta$ is the cosmological constant,
$\varphi$ is an scalar auxiliary field and $\alpha$ is the renormalized matter
fields central charge. We proceed calculating the Hamiltonian structure for the
model using light-cone variables, and after adding a suitable surface term to
(1) we end up with a Hamiltonian system of four first class constraints
$$\eqalignno{
\pi^{--} &={ \partial L\over {\partial (\partial _-g_{--} ) }}= 0= \Gamma ^1
\quad ,&(2a)\cr
\pi^{-+} &={\partial L\over {\partial (\partial _-g_{-+} )}}= 0 = \Gamma^2
\quad ,&(2b)\cr }$$

$$\eqalignno{
\phi _1&=\!{1\over 2}\!\!\left[\!(\qaz +\wsx )^2\!-\!{4\over {\alpha
^2}}(g_{++}
\pi ^{++})^2\!-\!
{4\over {\alpha }}(g_{++}\pi ^{++})\pi\!-\!{\alpha \qaz +g_{++}\qaz
+\wsx \over {
g_{++}}}\!+\!2\alpha \partial_+^2\wsx \!+\!\alpha ^2\beta g_{++}\right]\, ,&
(3a)\cr
\phi _2&=\pi\qaz +\wsx -2g_{++}\pi^{++}-\pi ^{++}\qaz +g_{++} \quad .
&(3b)\cr}$$
Equations (2) are the primary constraints while  equations (3) are secondary.
At this level the model reveals the well-known $SL(2,R)$ structure\ref{6} in a
very clear way, when we construct the following set of variables
$$
\eqalignno{
J^+ &={1\over {g_{++}}}(\phi_2-\phi_1)+{\alpha^2\beta\over {2}}\quad ,&(4a)\cr
 J^0&=j^0-x^-J^+\quad ,&(4b)\cr
j^0&=\left[ g_{++}\bigl(\pi^{++}+{\alpha \over 2}{\partial _+\varphi\over
g_{++}}\bigr) +{\alpha \over 2}\bigl(\pi-{\alpha \over 2}{\partial _+g_{++}
\over g_{++}}-\partial _+\varphi\bigr)\right]\quad ,&(4c) \cr
J^-&=j^--2x^-J^0-(x^-)^2J^+\quad ,&(4d)\cr
j^-&=\alpha ^2(g_{++}+1)\quad ,&(4e)\cr
b&=\pi-{\alpha \over {2}}{\qaz + g_{++}\over {g_{++}}}+\partial _+\varphi
\quad , &(4f) \cr }$$
which classically satisfy the Poisson bracket $SL(2,R)$ algebra
 $$\{J^a(x), J^b(y)\} =-2\epsilon ^{abc}\eta_{cd}J^d(x)\delta (x-y) +
\alpha ^2\eta ^{ab}\qaz +\delta (x-y)\quad .\eqno(5)$$

Using this structure it is also possible to calculate the quantum BRST charge
in a gauge independent way, expanding the energy momentum tensor in terms of
Virasoro modes;we define the Virasoro constraint as
$$\tau =T_m+T_S\approx 0 \eqno(6a)$$
where the $b$-field and gravity energy momentum tensor are respectively
$$
\eqalignno {
T_m&={1\over 4}b^2+{\alpha \over {2}}\partial _+b\quad ,&(6b)\cr
T_S&={1\over {2\alpha^2}}\eta_{ab}J^{a}J^{b}-\partial _+{J^0}\quad .&(6c)\cr
}$$

Imposing now the nilpotency $(\hat Q^2=0)$  for the BRST charge
$$\hat Q=c_0(L_0-a)+\sum _{n\neq 0}:c_nL_{-n}:
-{1\over 2}\sum ^{\infty}_{-\infty}
(m-n):c_{-m}c_{-n}b_{m+n}: \quad ,\eqno (7)$$
we obtain the usual relation for the central charges\ref{1,2} in a gauge
independent way
$$
c_{mat}+{3k\over{k+2}}-6k-28=0\quad ,\quad k ={\alpha^2\over 8}\quad .\eqno(8)
$$

As we mentioned earlier, the next step in our work was to fix completely the
gauge freedom for the model using the canonical formalism. This is a difficult
task because the theory is diffeomorphism invariant; as a consequence the
Hamiltonian is a linear combination of constraints
$$H_c=4\left[-{\sqrt {-g}\over {g_{++}}}\phi _1+{g_{-+}\over {g_{++}}}\phi _2
\right]\quad .\eqno (9)$$

In order to avoid this problem we use time dependent gauge fixing constraints.
Our result shows that this procedure annihilates all the physical degrees of
freedom for the pure gravity theory. However if we consider the model coupled
to matter fields
$$S={1\over 2}\int d^2x\sqrt {-g}[-\wsx\dal \wsx -X\dal X-\alpha
R\wsx -\gamma RX+(\gamma ^2+\alpha ^2)\beta]\quad ,\eqno (10)$$
It is possible to find a non-trivial reduced phase space leaving the canonical
procedure as a possible technique for quantization in this case.
\vskip 2cm
\penalty-200
\centerline {\bf Canonical Quantization of 2-D gravity coupled to matter
fields}
\vskip 1cm
\nobreak

We start adding a suitable surface term to the Lagrangian (10) in order to
find simpler expressions for the canonical momenta\newfoot{$^{(1)}$}{In this
case we work with Minkowiskian variables $x^0$ representing the time
variable.}; we find, after some algebra, the Lagrangian
$$
\eqalign{
{L} &={1\over 2\sqrt {-g}}\left[ (-g_{11}
\dot \varphi ^2+2g_{01}\dot \varphi \varphi '-g_{00}{\varphi '}^2)+\alpha
(\dot g_{11} \dot \varphi - 2g'_{01}\dot \varphi +g'_{00}\varphi ')
\right.\cr
& \left.+\alpha {g_{01}\over g_{11}}(g'_{11}\dot \varphi -\dot
g_{11}\varphi ')-\alpha ^2\beta  g\right]\cr
\cr & +{1\over 2\sqrt {-g}}\left[ (-g_{11}
\dot X ^2+2g_{01}\dot X X'-g_{00}{X'}^2)+\gamma
(\dot g_{11}\dot X - 2g'_{01}\dot X +g'_{00}X ')
\right.\cr
&\left.+\gamma {g_{01}\over g_{11}}(g'_{11}\dot X -\dot
g_{11}X')-\gamma ^2\beta g\right]\quad .\cr}\eqno (11)$$

The canonical momemta are derived from above, and read
$$\eqalignno{
\pi^{00}&={\partial L\over \partial \dot g_{00}}= 0 =\Gamma_1 &(12)\cr
\pi^{01}&={\partial L\over \partial \dot g_{01}}= 0 =\Gamma_2 &(13)\cr
\pi_\varphi&={1\over \sqrt{-g}}(g_{01}\varphi '-g_{11}\dot
\varphi)+{\alpha \over 2\sqrt{-g}}\left( \dot g_{11}-2g'_{01}+{g_{01}\over
g_{11}}g'_{11}\right) &(14)\cr
\pi^{11}&={\alpha\over 2\sqrt{-g}}\left( \dot\varphi-{g_{01}\over g_{11}}
\varphi '\right)+{\gamma \over {2\sqrt {-g}}}\left(\dot X-{g_{01}
\over {g_{11}}}X'\right)&(15)
\cr \bar \pi &={1\over \sqrt{-g}}(g_{01}X'-g_{11}\dot
X)+{\gamma \over 2\sqrt{-g}}\left( \dot g_{11}-2g'_{01}+{g_{01}\over
g_{11}}g'_{11}\right) \quad ,&(16)\cr }$$
where $\Gamma_1$ and $\Gamma_2$ are two primary constraints. As in the pure
case we find two secondary constraints $(\phi_1,\phi_2)$ which are also first
class
$$\eqalignno
{\phi _1&=-{1\over 2}\left( {\pi ^2}+ {\bar \pi ^2}\right)+
{(\alpha \pi +\gamma \bar \pi +2\pi^{11}g_{11})^2
\over {2(\alpha ^2+\gamma ^2)}}
-{(\alpha ^2+\gamma ^2)\beta g_{11}\over {2}}\cr
&\quad -{1\over 2}({\wsx '}^2+{X'}^2)+{g_{11}'\over{2g_{11}}}(\alpha\wsx '
+\gamma X')-\alpha \wsx ''-\gamma X''&(17a)\cr
\phi _2&=\wsx '\pi +X'\bar \pi -2g_{11}
\pi^{{11}'}-{g_{11} }'\pi ^{11}\qquad .&(17b)\cr }$$

The diffeomorphism invariance of the model is verified through the constraint
algebra:
$$\eqalignno{
\lbrace \phi _1 (x),\phi _1(y)\rbrace &=[\phi _2(x)+\phi _2(y)]\qaz x
\delta  (x-y) &(18a)\cr
\lbrace \phi _1 (x),\phi _2(y)\rbrace &=[\phi _1(x)+\phi _1(y)]\qaz x
\delta (x-y) &(18b)\cr
\lbrace \phi _2 (x),\phi _2(y)\rbrace &=[\phi _2(x)+\phi _2(y)]\qaz x
\delta (x-y)\quad . &(18c)\cr}$$

This invariance leads us to impose a set of gauge fixing constraints, following
the Dirac method. To fix the gauge freedom completely we need four of them:
$$\eqalignno{
\Gamma_3&= g_{11} - g_{00} +2 =0 & (19a)\cr
\Gamma_4&=g_{11} - g_{01} + 1 = 0 & (19b)\cr
\phi _3&=g_{11}-F(t,x)&(20a)\cr
\phi_4&={2\over{\alpha^2+\gamma^2}}(\alpha\pi+\gamma\bar\pi)\quad ,&(20b)\cr}$$
which transform the whole set of constraints $(\Gamma_1,\Gamma_2,\Gamma_3,
\Gamma_ 4,\phi_1,\cdots\phi_4)$ into a second class system. The calculation of
the Dirac brackets, which gives us the structure of the reduced phase space,
can be made in a two step procedure. In the first step we eliminate the
spurious
degrees of freedom corresponding to the set  $\{(\Gamma_1, \Gamma_2, \Gamma_3,
\Gamma_ 4)\}$. While considering the second step, it is simpler to work with an
extended Hamiltonian (which includes the linear combination of all constraints
with arbitrary coefficients)
$$H_e=-\left(v_1-{\sqrt {-g}\over g_{11}}\right)\phi _1+\left(v_2+{g_{01}
\over g_{11}}\right)\phi _2\quad ,\eqno (21)$$
in order to stop the generation of new constraints coming from the time
consistency of $\phi_3$ and $\phi_4$. This procedure in fact fixes the unknown
coefficients  $v_1$ and $v_2$.

In order to construct the Dirac matrix
$$\Delta _{ij}(x,y)=\lbrace \phi _i (x),\phi _j(y)\rbrace \quad , \eqno (22)$$
we calculate the Poisson brackets between the constraints. The non-zero ones
are in this case:
$$\eqalignno {
\lbrace \phi _1(x),\phi _3(y)\rbrace &\equiv \Delta_{13}(x,y)
=-{2g_{11}^2\pi ^{11}(x)\over {\rho ^2}}\delta(x-y)&(23a)\cr
\lbrace \phi _1(x),\phi _4(y)\rbrace &\equiv \Delta_{14}(x,y)
=\!\!\left( {g_{11}'\over {g_{11}}}-
{2\over {\rho ^2}} \!\left[\alpha \wsx '+\gamma X'\right]\!\!\right)\!\!(x)
\partial _{x}\delta (x\!-\!y)-2\partial ^2_x \delta (x-y) &(23b)\cr
\lbrace \phi _2(x),\phi _3(y)\rbrace &\equiv \Delta_{23}(x,y) =2g_{11}(x)
\partial _x\delta (x-y)+g_{11}'(x)\delta (x-y)\quad . &(23c)\cr } $$

The Dirac brackets require the calculation of the inverse of this matrix. After
a tedious calculation we find for the non zero elements
$$\eqalignno{
\Delta ^{-1}_{32}(x,y)&={\epsilon (x-y)\over {4\sqrt {g_{11}(x) g_{11}(y)}}
} &(24a)\cr
\Delta ^{-1}_{41}(x,y)&=\left\{ \int ^y_{-\infty }  dz
{\epsilon (x-z)\over {4}}g_{11}(z)e^{-{2\over {\rho ^2}}(\alpha
\wsx (z)+\gamma X(z))}\right\} e^{{2\over {\rho ^2}}(\alpha \wsx (y)
+\gamma X(y))}g_{11}(y) &(24b) \cr
\Delta  ^{-1} _{42}(x,y)&=-{2\over {\rho ^2}}\left[\int ^y_{-\infty }
dz g_{11}\pi ^{11}(x)\Delta _{41}^{-1}
(x,z)\sqrt {g_{11}(z)} {1\over {\sqrt {g_{11}(y)}}}\right]\quad .&(24c)\cr}$$
Thus we find the following Dirac brackets:

$$\eqalign {
\lbrace \wsx (x),\wsx (y)&\rbrace _D=-{2\alpha \over {\rho ^2}}
\left(2g_{11}\pi ^{11}(x){\alpha \over {\rho ^2}}-\pi (x)\right)
\int ^x_{-\infty }  dz{\epsilon (z-y)\over {4}} C^-(z) C^+(x)
{\sqrt {g_{11}(z)}\over{\sqrt {g_{11}(x)}}}\cr
&+{2\alpha \over {
\rho ^2}}\left(g_{11}\pi ^{11}(y){\alpha \over {\rho ^{2}}}
-\pi (y)\right) \int ^y_{-\infty }  dz
{\epsilon (x-z)\over {4}}C^-(z)C^+(y){\sqrt {g_{11}(z)}\over
{\sqrt {g_{11}(y)}}} \cr
&-{2\alpha \over {\rho ^2}}\left\{ \wsx '(y)\left[\int ^y_{-\infty } dz
{g_{11}\pi ^{11}(z)\over {\rho ^2}}
\int ^x_{-\infty }dw {\epsilon (w-z)\over {4}}C^-(w)
C^+(z){\sqrt {g_{11}(w)} \over {\sqrt {g_{11}(y)}}}\right] \right. \cr
&-\left.
 \wsx '(x)\!\!\left[\int ^x_{-\infty } dz {g_{11}\pi ^{11}(z)\over {\rho ^2}}
\!\!\int ^y_{-\infty }dw {\epsilon (w-z)\over {4}}C^-(w)
C^+(z){\sqrt {g_{11}(w)}\over {\sqrt {g_{11}(x)}}}\right]\!\right\}\quad ,\cr}
\eqno(25)$$
where
$$C^{\pm }(z)=e^{{\pm }{1\over {\rho ^2}}(\alpha \wsx (z)+\gamma X(z))}\quad .
\eqno (26)$$

Further Dirac brackets can be computed in a straightforward but tedious way; we
find for the purely matter field Dirac bracket the expression:
$$\eqalign {
\lbrace X (x),X(y)&\rbrace _D\!=\!-{2\gamma \over {
\rho ^2}}\!\!\left(\!\!2g_{11}\pi ^{11}(x)
{\gamma \over {\rho ^2}}\!-\!\bar \pi (x)\!\!\right)\!\!
\int ^x_{-\infty } \!\!\!\! dz
{\epsilon (z-y)\over {4}}\sqrt {g_{11}(z)} C^-(z) C^+(x){1\over
{\sqrt {g_{11}(x)}}}\cr &+{2\gamma \over {
\rho ^2}}\left(2g_{11}\pi ^{11}(y){\gamma \over {\rho ^{2}}}
-\bar \pi (y)\right) \int ^y_{-\infty }  dz
{\epsilon (x-z)\over {4}}\sqrt {g_{11}(z)}C^-(z)C^+(y){1\over
{\sqrt {g_{11}(y)}}} \cr &-
{2\gamma \over {\rho ^2}}\left\{ X'(y)\left[\int ^y_{-\infty } dz {g_{11}
\pi ^{11}(z)\over {\rho ^2}}
\int ^x_{-\infty }dw {\epsilon (w-z)\over {4}}C^-(w)
C^+(z){\sqrt {g_{11}(w)} \over {\sqrt {g_{11}(y)}}}\right] \right. \cr &-\left.
 X'(x)\left[\int ^x_{-\infty } dz {g_{11}\pi ^{11}(z)\over {\rho ^2}}
\int ^y_{-\infty }dw {\epsilon (w-z)\over {4}}C^-(w)
C^+(z){\sqrt {g_{11}(w)} \over {\sqrt {g_{11}(x)}}}\right] \right\}
\quad .\cr }\eqno(27)$$
 \vskip .2cm
The mixed gravity-matter fundamental bracket is
$$\eqalign
{\lbrace \wsx (x),X(y)&\rbrace _D\!=\!-{2\gamma \over {
\rho ^2}}\!\left(\!2g_{11}\pi ^{11}(x)
{\alpha \over {\rho ^2}}\!-\!\pi (x)\!\!\right)\!\!\! \int ^x_{-\infty }
\!\!\!\!  dz
{\epsilon (z\!-\!y)\over {4}}\sqrt {g_{11}(z)} C^-(z) C^+(x){1\over
{\sqrt {g_{11}(x)}}}\cr &+{2\gamma \over {
\rho ^2}}\left(g_{11}\pi ^{11}(y){\alpha \over {\rho ^{2}}}
-\pi (y)\right) \int ^y_{-\infty }  dz
{\epsilon (x-z)\over {4}}\sqrt {g_{11}(z)}C^-(z)C^+(y){1\over
{\sqrt {g_{11}(y)}}} \cr &-
{2\gamma \over {\rho ^2}}\left\{ \wsx '(y)\left[\int ^y_{-\infty } dz
{g_{11}\pi ^{11}(z)\over {\rho ^2}}
\int ^x_{-\infty }dw {\epsilon (w-z)\over {4}}C^-(w)
C^+(z){\sqrt {g_{11}(w)}\over {\sqrt {g_{11}(y)}}}\right] \right. \cr &-\left.
 \wsx '(x)\left[\int ^x_{-\infty } dz {g_{11}\pi ^{11}(z)\over {\rho ^2}}
\int ^y_{-\infty }dw {\epsilon (w-z)\over {4}}C^-(w)
C^+(z){\sqrt {g_{11}(w)}\over {\sqrt {g_{11}(x)}}}\right] \right\}
\quad .\cr }\eqno (28)$$
Furthermore the brackets for the fields $\varphi$, $X$ and their momenta
\vskip .4cm
$$\eqalign {\lbrace \pi (x),\wsx (y)& \rbrace _D=-\delta (x-y)+
{2\alpha \over {\rho ^2}}\biggl[
\partial _y \Bigl[\left(\int ^x_{-\infty }  dz
{\epsilon (z-y)\over {4}}\sqrt {g_{11}(z)} C^-(z) C^+(x){1\over {\sqrt {
g_{11}(x)}}}\right) \cr
&\times\left({\alpha g_{11}'
(z)\over {2g_{11}(z)}}-\wsx '(z)\right)\Bigr]\biggr]
+{2\alpha ^2\over {\rho ^2}} \qaz y^2\bigl[\int ^x_{-\infty }  dz
{\epsilon (z-y)\over {4}}{\sqrt {g_{11}(z)}\over {\sqrt {g_{11}(x)}}}
C^-(z) C^+(x)\bigr]\cr
&+{2\alpha \over {\rho ^2}}\qaz y\left[\pi (y)[\int ^x_{-\infty } dz {g_{11}
\pi ^{11}(z)\over{\rho ^2}}
\int ^y_{-\infty }dw {\epsilon (w-z)\over {4}}C^-(w)
C^+(z){\sqrt {g_{11}(w)}\over {\sqrt {g_{11}(x)}}}]\right]\quad , \cr }\eqno
(29)$$
\vskip .2cm

$$
\eqalign {\lbrace \bar \pi (x),X (y)& \rbrace _D=-\delta (x-y)+{2\gamma
\over {\rho ^2}}\biggl[
\partial _y \Bigl[\left(\int ^x_{-\infty }  dz
{\epsilon (z-y)\over {4}}\sqrt {g_{11}(z)} C^-(z) C^+(x){1\over {
\sqrt {g_{11}(x)}}}\right) \cr
& \times\left({\gamma g_{11}'
(z)\over {2g_{11}(z)}}-X '(z)\right)\Bigr]\biggr]
+{2\gamma ^2\over {\rho ^2}} \qaz y^2\bigl[\int ^x_{-\infty }  dz
{\epsilon (z-y)\over {4}}{\sqrt {g_{11}(z)}\over {\sqrt {
g_{11}(x)}}} C^-(z) C^+(x)\bigr]\cr
&+{2\gamma \over {\rho ^2}}\qaz y\left[\bar \pi (y)[\int ^x_{-\infty } dz
{g_{11}\pi ^{11}(z)\over {\rho ^2}}
\int ^y_{-\infty }dw {\epsilon (w-z)\over {4}}C^-(w)
C^+(z){\sqrt {g_{11}(w)}\over {\sqrt {g_{11}(x)}}}]\right];\cr }\eqno (30)$$
\vskip .2cm
we have also
\vskip .2cm
$$
\eqalign {\lbrace  \pi (x),X (y)& \rbrace _D={2\gamma
\over {\rho ^2}}\biggl[
\partial _y \Bigl[\left(\int ^x_{-\infty }  dz
{\epsilon (z-y)\over {4}}\sqrt {g_{11}(z)} C^-(z) C^+(x){1\over {\sqrt {g_{11}
(x)}}}\right)\times \cr &\left({\alpha g_{11}'
(z)\over {2g_{11}(z)}}-\wsx '(z)\right)\Bigr]\biggr]
+{2\gamma \over {\rho ^2}}\alpha \qaz y^2\bigl[\int ^x_{-\infty }  dz
{\epsilon (z-y)\over {4}}{\sqrt {g_{11}(z)}\over {
\sqrt {g_{11}(x)}}} C^-(z) C^+(x)
\bigr]\cr
&+{2 \gamma \over {\rho ^2}}\qaz y\left[ \pi (y)[\int ^x_{-\infty } dz
{g_{11}\pi ^{11}(z)\over{\rho ^2}}
\int ^y_{-\infty }dw {\epsilon (w-z)\over {4}}C^-(w)
C^+(z){\sqrt {g_{11}(w)}\over {\sqrt {g_{11}(x)}}}]\right],\cr }\eqno (31)$$
and
$$
\eqalign {\lbrace \bar \pi (x),\wsx (y)& \rbrace _D={2\alpha
\over {\rho ^2}}\biggl[\partial _y \Bigl[\left(\int ^x_{-\infty }  dz
{\epsilon (z-y)\over {4}}\sqrt {g_{11}(z)} C^-(z) C^+(x){1\over {
\sqrt {g_{11}(x)}}}\right) \times \cr
&\left({\gamma g_{11}'
(z)\over {2g_{11}(z)}}\!-\!X '(z)\!\right)\!\Bigr]\!\biggr]\!
+\!{2\alpha \over {\rho ^2}}\gamma \qaz y^2\bigl[\!\int ^x_{-\infty }\!\!\! dz
{\epsilon (z-y)\over {4}}\sqrt {g_{11}(z)} C^-(z) C^+(x){1\over {
\sqrt {g_{11}(x)}}}\bigr]\cr
&+{2\alpha \over {\rho ^2}}\qaz y\left[\bar \pi (y)[\int ^x_{-\infty } dz
{g_{11}\pi ^{11}(z)\over {\rho ^2}}
\int ^y_{-\infty }dw {\epsilon (w-z)\over {4}}C^-(w)
C^+(z){\sqrt {g_{11}(w)} \over {\sqrt {g_{11}(x)}}}]\right].\cr }\eqno (32)$$
\vskip .3cm

For brackets involving  $\pi ^{11}$ we arrive at \vskip .2cm
$$\eqalign {
\lbrace \wsx (x),\pi ^{11}(y)&\rbrace _D=-{\varphi (x)\epsilon (x-y)
\over {4\sqrt {g_{11}(x) g_{11}(y)}}}\cr
&-{2\alpha \over {\rho ^2}}\left(\int ^y_{-\infty }  dz
{\epsilon (x-z)\over {4}}\sqrt {g_{11}(z)}C^-(z)C^+(y){1\over {
\sqrt {g_{11}(y)}}}({2g_{11}{\pi ^{11}}^2(y)\over
{\rho ^2}}\right.\cr &-{\rho ^2\beta \over {2}}-\left.{(\alpha \wsx '
+\gamma X')\over {2}}{g_{11}'(y)\over {g_{11}^2(y)}})\right)\cr
&-{2\alpha \over {\rho ^2}}\partial _y\left[\int ^y_{-\infty }  dz
{\epsilon (x-z)\over {4}}\sqrt {g_{11}(z)}C^-(z)C^+(y){1\over
{\sqrt {g_{11}(y)}}}{(\alpha \wsx '+\gamma
X')(y)\over {g_{11}(y)}}\right]\cr
&+{4\alpha \over {\rho ^2}}{\pi ^{11}}'(y)\left[ \int ^x_{-\infty } dz
{g_{11}\pi ^{11}(z)\over{\rho ^2}}
\int ^y_{-\infty }dw {\epsilon (w-z)\over {4}}C^-(w)
C^+(z){\sqrt {g_{11}(w)} \over {\sqrt {g_{11}(x)}}}\right]\cr
&-{2\alpha \over {\rho ^2}}\qaz y\!\left[\pi ^{11}(y)\bigl[ \!\int ^x_{-\infty}
\!\!\! dz {g_{11}\pi ^{11}(z)\over {\rho ^2}}\!
\int ^y_{-\infty }\!\!\!dw {\epsilon (w-z)\over {4}}C^-(w)
C^+(z){\sqrt {g_{11}(w)}\over \sqrt {g_{11}(x)}}\bigr]\!\right]\cr }$$
{\hfill(33)}
and
$$\eqalign {\lbrace X (x),\pi ^{11}(y)&\rbrace _D=-{X (x)\epsilon (x-y)
\over {4\sqrt {g_{11}(x) g_{11}(y)}}}\cr
&-{2\gamma \over {\rho ^2}}\left(\int ^y_{-\infty }  dz
{\epsilon (x-z)\over {4}}\sqrt {g_{11}(z)}C^-(z)C^+(y){1\over {
\sqrt {g_{11}(y)}}}({2g_{11}{\pi ^{11}}^2(y)\over
{\rho ^2}}\right.\cr &-{\rho ^2\beta \over {2}}-\left.{(\alpha \wsx '
+\gamma X')\over {2}}{g_{11}'(y)\over {g_{11}^2(y)}})\right)\cr
&-{2\gamma \over {\rho ^2}}\partial _y\left[\int ^y_{-\infty }  dz
{\epsilon (x-z)\over {4}}\sqrt {g_{11}(z)}C^-(z)C^+(y){1\over
{\sqrt {g_{11}(y)}}}{(\alpha \wsx '+\gamma
X')(y)\over {g_{11}(y)}}\right]\cr
&+{4\gamma \over {\rho ^2}}{\pi ^{11}}'(y)\left[ \int ^x_{-\infty } dz
{g_{11}\pi ^{11}(z)\over{\rho ^2}}
\int ^y_{-\infty }dw {\epsilon (w-z)\over {4}}C^-(w)
C^+(z){\sqrt {g_{11}(w)}\over {\sqrt {g_{11}(x)}}}\right]\cr
&-{2\gamma \over {\rho ^2}}\qaz y\!\left[\pi ^{11}(y)\bigl[\!\int ^x_{-\infty }
\!\!\! dz {g_{11}\pi ^{11}(z)\over {\rho ^2}}\!
\int ^y_{-\infty }\!\!\!dw {\epsilon (w-z)\over {4}}C^-(w)
C^+(z){\sqrt {g_{11}(w)}\over {\sqrt {g_{11}(x)}}}\bigr]\!\right] .\cr }$$
{\hfill (34)}

\vskip 2cm
\penalty-500
\centerline {\bf Conclusions}
\vskip .5cm
\nobreak

We find a closed Dirac algebra for the set of physical fields of the theory.
There is  always a non-local tail in the commutator function, in such way that
the theory does not simplify in some limit. For $\alpha \to -\infty$
(semi-classical limit) the exponential functions simplify to one;  the
commutators go to zero, with a ${\cal O}(1/\alpha)$ non-local correction of the
type
$$\int _{-\infty}^x dx \epsilon (z-y)g_{11}(z) g_{11}(x)\quad .\eqno(35)$$

In this example, there is a feature that can not be forgotten, concerning well
known non-perturbative results obtained for several correlation functions.
Simple results concerning correlators for vertex functions, cannot be obtained
in this case. If we delete the matter field $X$, the canonical quantization
turns out to imply  a trivial result, whereas the exact solution is far from
trivial\ref{1,3}. This single result suggests that canonical quantization
of higher dimensional gravity may be missing an entire sector of the theory. A
second example concerning the impossibility of describing non-perturbative
effects by means of canonical quantization concerns $3-D$ topological Yang
Mills theory, where canonical quantization leads to results far simpler than
the expected ones.

Finally, the non-local tail can be traced back to the smearing of the
fundamental fields as implied by the gravity ``fog", and should play an
important role in problems tied with the horizon in quantum gravity.
\vskip 1cm
\noindent {\bf Acknowledgement:} This work of E.A. and M.C.B.A. has been
partially supported by CNPq, while that of F.P. Devecchi has been supported by
FAPESP.

\vskip .5cm
\centerline {\bf References}
\vskip .5cm
\refer[[1]/A. Polyakov, Mod. Phys. Lett. {\bf A}2 (1987)893;]

\refer[/V. Knizhnik, A. Polyakov and A. Zamolodchikov, Mod. Phys. Lett {\bf A}3
(1988)819.]

\refer[[2]/F. David, Mod. Phys. Lett.{\bf A}3 (1988)165;]

\refer[/J. Distler and H. Kawai, Nucl. Phys. {\bf B}321 (1988)509.]

\refer[[3]/G. Moore, N. Seiberg and M. Staudacher, Nucl. Phys. {\bf B}362
(1991)665.]

\refer[[4]/M. Goulian and M. Li, Phys. Rev. Lett. {\bf 66} (1991)2051;]

\refer[/P. di Francesco and D. Kutasov, Nucl. Phys. {\bf B}375 (1992)119;]

\refer[/E. Abdalla, M.C.B. Abdalla, D. Dalmazi and K. Harada, Phys. Rev. Lett.
{\bf 68} (1992)1641;]

\refer[/K. Aoki, E. D'Hoker  Mod. Phys. Lett.  {\bf A}7
(1992)333.]

\refer[[5]/E. Abdalla, M.C.B. Abdalla and K. Rothe, {\it Non peturbative
methods
in two dimension quantum fields theory}, World Scientific Publishing, 1990.]

\refer[[6]/E. Abdalla, M.C.B. Abdalla, J. Gamboa and A. Zadra, Phys. Lett.
{\bf B}273 (1991)222.]

\refer[[7]/A. Mikovic, Queen Mary College preprint-91/22;]

\refer[/E. Egorian, and R. Manvelian, Mod. Phys. Lett. {\bf A}5 (1990)2371.]

\end